\RequirePackage[l2tabu, orthodox]{nag}

\documentclass[fontsize=12pt,a4paper,numbers=enddot,parskip=half]{scrartcl} 

\usepackage{float}

\usepackage[T1]{fontenc}
\usepackage{lmodern} 

\usepackage{amsfonts}
\usepackage{amsmath}
\usepackage{amsthm}
\usepackage{amssymb}
\usepackage{eurosym}
\usepackage{bbm}

\setkomafont{section}{\rmfamily\fontseries{b}\selectfont}
\setkomafont{subsection}{\rmfamily\itshape\mdseries}
\setkomafont{subsubsection}{\rmfamily\itshape\mdseries}
\setkomafont{paragraph}{\rmfamily\itshape\mdseries}
\setkomafont{title}{\rmfamily\large\fontseries{b}\selectfont}
\setkomafont{captionlabel}{\rmfamily\large\fontseries{b}\small\selectfont}

\usepackage{multirow}
\usepackage[font={small},justification=raggedright,singlelinecheck=false]{caption}
\newcommand{\mytablefont}{\vspace{2mm}\fontsize{9}{10.8} \selectfont}

\usepackage{tikz}
\newcommand{\xShiftModelTikz}{1.3cm}
\newcommand{\xShiftModelTikzDense}{0.5cm}

\usetikzlibrary{positioning,arrows}

\usepackage[caption=false]{subfig}
\usepackage{todonotes}

\usepackage[ansinew]{inputenc}
\usepackage[english]{babel}

\begin{hyphenrules}{english}
\end{hyphenrules}

\usepackage[top=3cm, bottom=3cm, left=3cm, right=3cm]{geometry}
\usepackage{setspace}
\usepackage{rotating} 
\usepackage{dcolumn}
\newcolumntype{d}[1]{D{.}{.}{#1}}

\usepackage{graphicx}
\usepackage{booktabs}
\usepackage{epstopdf}
\usepackage{enumerate}
\usepackage[hang]{footmisc}
\usepackage[longnamesfirst]{natbib}
\usepackage{authblk}
\usepackage{colortbl}
\usepackage{placeins}
\usepackage{tabularx}
\usepackage{makecell}
\usepackage{wrapfig}
\newcolumntype{Y}{>{\centering\arraybackslash}X}
\newcolumntype{R}{>{\flushright\arraybackslash}X}
\newcolumntype{L}[1]{>{\raggedright\let\newline\\\arraybackslash\hspace{0pt}}m{#1}}

\author{\vspace{5mm} \small MORITZ SCHERRMANN\thanks{Institute for Finance \& Banking, Ludwig-Maximilians-Universit\"at M\"unchen, Ludwigstr.\ 28 RB, 80539 Munich, Germany. E-mail: scherrmann@lmu.de}, RALF ELSAS\thanks{Institute for Finance \& Banking, Ludwig-Maximilians-Universit\"at M\"unchen, Ludwigstr.\ 28 RB, 80539 Munich, Germany. E-mail: elsas@lmu.de}}
\title{\Large Earnings Prediction Using Recurrent Neural Networks}
\date{}

\begin{document}

\maketitle
\vspace{2cm}
{\rmfamily\fontseries{b}\selectfont Abstract}

\smallskip
\small \singlespacing 
Firm disclosures about future prospects are crucial for corporate valuation and compliance with global regulations, such as the EU's MAR and the US's SEC Rule 10b-5 and RegFD. To comply with disclosure obligations, issuers must identify nonpublic information with potential material impact on security prices as only new, relevant and unexpected information materially affects prices in efficient markets. Financial analysts, assumed to represent public knowledge on firms' earnings prospects, face limitations in offering comprehensive coverage and unbiased estimates. This study develops a neural network to forecast future firm earnings, using four decades of financial data, addressing analysts' coverage gaps and potentially revealing hidden insights. The model avoids selectivity and survivorship biases as it allows for missing data. Furthermore, the model is able to produce both fiscal-year-end and quarterly earnings predictions. Its performance surpasses benchmark models from the academic literature by a wide margin and outperforms analysts' forecasts for fiscal-year-end earnings predictions.
\normalsize
\vspace{2cm}

\newpage
\onehalfspacing

\section{Introduction}

Firm disclosures regarding future prospects are critical for determining corporate valuation, necessitating global regulations such as the European Union's Market Abuse Regulation (MAR) and the United States' Securities and Exchange Commission's Rule 10b-5 and Regulation Fair Disclosure (RegFD). These regulations mandate the accurate and timely disclosure of material nonpublic information. However, issuers encounter difficulties in identifying nonpublic information that could impact security prices, as only new, relevant and unexpected information causes price fluctuations in efficient markets. Consequently, a benchmark for "new" news is essential for issuers to comply with disclosure obligations.

In this context, financial analysts are often assumed to represent public knowledge concerning firms' earnings prospects. The German security markets supervisory authority (BaFin) recently even designated analysts' predictions as their regulatory measure for capital market expectations on firms' earnings. Given financial analysts' important role in advising investors and providing insights into earnings prospects through the publication of estimates on future earnings and related key ratios, such a "benchmark" initially appears reasonable.

However, reliance on analysts has certain limitations. Firstly, their estimates may be subject to biases arising from conflicts of interest and existing research shows limited predictive power in explaining stock price reactions to earnings disclosures \cite[see][]{abarbanell1991analysts,o1988analysts,bradshaw2012re}. Secondly, analyst coverage is limited, particularly outside the US.\footnote{For example, in Germany in the period from 1998 to 2022, less than 50\% of CDAX companies received coverage for fiscal-year-end earnings and less than 20\% for quarterly earnings.} Moreover, recent regulatory compliance burdens have prompted analyst firms and financial intermediaries to further reduce coverage, especially for smaller, less prominent companies.

In this study, we develop a neural network to forecast future firm earnings, utilizing approximately four decades of financial data. Machine learning models can effectively process and analyze large amounts of data to predict corporate earnings. By employing advanced algorithms, these models can quickly sift through historical data, financial statements and market trends to generate accurate forecasts. This automation potentially addresses the gaps in analysts' coverage and the model may even reveal hidden insights by decoding complex patterns in the input data. Consequently, such an earnings prediction model provides a more comprehensive benchmark for the capital market's expectation concerning firms' earnings prospects, thereby assisting issuers in better adhering to disclosure regulations and improving guidance for investors' investment decisions.

Key features that differentiate our model from previous statistical approaches in the literature are that we avoid selectivity biases through high input data demand, as we strive for a parsimonious specification and explicitly allow for missing data in the time-series of accounting data used as predictors. This approach ensures that our model is not biased towards firms with more complete or longer available data, thus avoiding survivorship biases, which is a common issue in previous studies. Also different from most of the academic literature, we provide forecast results not just for fiscal-year-end predictions but also for quarterly earnings. This allows us to capture and study the dynamics of earnings over shorter time periods and is particularly useful for issuers and investors, as quarterly earnings announcements are the most important source of information that determines stock prices.   

The structure of our study is organized as follows. In Section \ref{paper2_sec_literature_review}, we provide a brief overview of the existing literature on earnings prediction and discuss how our study contributes to this field. Section \ref{paper2_sec_methodology} describes the design of the neural network used for earnings prediction, the training procedure and the benchmarks used to validate the model's predictive power. We emphasize that the model's performance is always measured using out-of-sample data not used in the model's training. In Section \ref{paper2_sec_data}, we describe the data used in our study, including its selection and provide descriptive statistics. 

Section \ref{paper2_sec_results} contains the empirical results. We report the performance of our model as well as benchmark models, including a random walk as a time-series model, a cross-sectional model by \cite{hess2019incorporating} and analysts' forecasts. We report performance results for both end-of-fiscal-year and quarterly earnings predictions, for the overall period of analysis as well as subsamples. These subsamples differentiate between predictive performance for different firm sizes, industry roots of firms and the time-period before, during and after the COVID-19 pandemic. This last conditioned subsample allows us to test model performance during a period characterized by a severe and unexpected (i.e., exogenous) shock to firm operations, as the COVID-19 pandemic led to shutdowns with regard to operations, the rise of home-office-work, associated changes in consumer demand and temporary financial squeezes. Lastly, we report results for a matched-firm approach that allows us to test model performance while controlling for the selectivity introduced by running all previous tests on the subsample of firms with analyst coverage. Section \ref{paper2_sec_conclusion} concludes our study.

In terms of results, our predictive model is at least at par with analysts' forecasts, but performance quite generally is higher for fiscal-year-end forecasts, also when conditioning on industries, firm size and the time-period of the COVID-19 pandemic. Used benchmark models are consistently outperformed by both our model as well as analysts' forecasts by a wide margin. 

\section{Literature Review}
\label{paper2_sec_literature_review}
We broadly divide the stream of literature that compares the accuracy\footnote{Accuracy in this context measures the difference between the predicted and actual earnings. We will use the Median Percentage Difference (MPD) and Median Absolute Percentage Difference (MAPD) in our analyses. See Section \ref{paper_2_sec_performance_measures}.} of model-based forecasts and analysts' forecasts by the used methodology, i.e.~cross-sectional and time-series models and discuss cross-sectional predictive models first. 

The literature stream of cross-sectional models starts with \cite{fama2000forecasting}, where the authors predict future annual profitability and earnings using year-by-year cross-sectional regressions between 1964 and 1996. They use the last annual earnings result, a firm's market value, total dividends as well as a dividend dummy to predict the next years earnings. \cite{hou2012implied} extend the idea of \cite{fama2000forecasting} by regressing future annual earnings on total assets, dividends, earnings and accruals. \cite{li2014evaluating} use in addition past earnings and book values instead of dividends as predictors. Despite the fact that all these modelling steps lead to improvements in terms of accuracy and earnings response coefficients\footnote{The Earnings Response Coefficient (ERC) is one measure of predictive model performance which we will also rely upon in our analyses. The ERC is calculated by regressing the stock price's abnormal return (the return that exceeds the expected return) on the unexpected earnings (the difference between actual earnings and expected earnings). The predictive model's earnings prediction is then used as the expected earning. A model performs better, all else equal, when it explains more of the abnormal returns on average.}, none of these models is able to outperform the predictive performance of analysts.

\cite{hess2019incorporating} contend that the enhanced performance of analysts relative to cross-sectional models can be attributed to their informational advantage, particularly as these studies often overlook quarterly data that analysts routinely utilize. \cite{hess2019incorporating} use the information of already reported quarterly results to predict the sum of all quarters that are not reported yet. Hence, the authors annual earnings prediction is the sum of reported and predicted quarters. By incorporating quarterly earnings results, \cite{hess2019incorporating} strongly improve forecast accuracy over short- and long-term horizons. However, analyst forecasts are still superior in terms of accuracy.\footnote{\cite{azevedo2021earnings} use an annual cross-sectional model with analyst forecasts as predictors. Their model outperforms the above discussed benchmarks from the literature in terms of forecast accuracy, but relying on analyst forecast prohibits the model to serve as an alternative to compensate for low and further reducing coverage by analysts.}

There is rich literature on time-series models for earnings prediction between 1978 and 1995.\footnote{\footnotesize{See \cite{bradshaw2012re} for a detailed summary of time-series literature comparing time-series and analyst forecasts.}} The main conclusion of that literature is that analysts are better at predicting future earnings than time-series models \citep{bradshaw2012re}. In addition, the studies were criticized due to relying on only small sample sizes, comprising only large, established companies and due to their data intensity, see e.g. \cite{kothari2001capital}.

To identify a time-series model for benchmarking purposes, we discuss three recent studies. \cite{bradshaw2012re} use a random walk model to predict annual earnings per share (EPS) of US companies between 1983 and 2008. They find that the EPS forecasts of the simple random walk model are more accurate than analysts' forecasts over longer horizons, for smaller or younger firms, when analysts forecast are negative or if there were large changes in EPS.

\cite{ball2018automated} augment classical quarterly accounting variables with high-frequent market information as stock returns, stock volatility or oil prices. The authors find that their time-series forecasts are more accurate than analysts' and that combining time-series forecasts with analysts' forecasts systematically outperforms analysts alone. 

Similar to our analysis, \cite{elend2020earnings} use a machine learning model to predict future, quarterly earnings. More precisely, they use a long short-term memory neural network \citep{LSTM} processing a time-series of 20 periods with 19 accounting predictors from Compustat and 11 stock market predictors from CRSP. 

\cite{ball2018automated} and \cite{elend2020earnings} are two of the few studies which predict quarterly instead of annual earnings results. However, due to the large number of predictors required in both approaches, the samples contain mostly large and established firms, thereby likely inducing survivorship bias.

In summary, the previous literature on predicting firm earnings statistically is characterized by using a large number of predictor variables (both cross-sectionally and in the time-series), which is likely to introduce selection biases in the results. Also, most studies focus on predicting fiscal-year-end results, ignoring quarterly earnings (both as an input as well as the predicted variable). Our study addresses both issues by relying on cross-sectional and time-series data but allowing for missing data and heterogeneous time-series' length. The model thus mitigates selectivity and survivorship bias. In addition, our neural network will be trained to predict both quarterly and fiscal-year-end earnings. 

\FloatBarrier

\section{Methodology}
\label{paper2_sec_methodology}
\subsection{Model}
\label{paper2_sec_model}
We think that various types of variables contribute to the prediction of a company's future earnings. These range from accounting data available at a quarterly frequency to daily market data, as well as time-invariant factors such as the company's industry or the quarter in which an earnings announcement is made. Classical models like cross-sectional regressions or time series models are not able to handle different input types or they suffer from survivorship or success biases. We aim to create a model which is able to handle time-series inputs as well as time invariant variables at the same time.

Our model is a combination of a recurrent neural network (RNN) and three feed-forward layers. The recurrent neural network consists of two layers of gated recurrent units (GRUs), introduced by \cite{GRU}. The layers have 76 and 38 hidden neurons, respectively. We design the RNN to process five years of quarterly accounting data, resulting in a time series of 20 quarters. The RNN outputs a time series of 38 dimensional arrays of which we use only the last one, since this array contains all the information of its predecessors. For market data input, we avoid using its time series structure by considering only the changes in the corresponding variable within the most recent quarter. We also test a design which preserves the time series structure of the market data using a second RNN. However, this model is more complex and data intensive without outperforming the approach using the reduced market data.

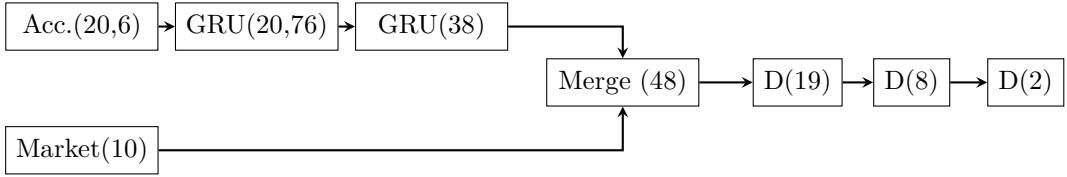
\begin{figure}

\begin{tikzpicture}
\tikzstyle{every node}=[font=\footnotesize]

\tikzstyle{input} = [rectangle, minimum width=2cm,  text centered, draw=black]
\tikzstyle{gru} = [rectangle, minimum width=2cm,  text centered, draw=black]
\tikzstyle{dense} = [rectangle, minimum width=1cm, text centered, draw=black]
\tikzstyle{merge} = [rectangle, minimum width=2cm,  text centered, draw=black]
\tikzstyle{arrow} = [thick,->,>=stealth]

\node (InputAccounting) [input,  xshift=\xShiftModelTikz] {Acc.(20,6)};
\node (Gru1Accounting) [gru, right of=InputAccounting, xshift=\xShiftModelTikz]
{GRU(20,76)};
\node (Gru2Accounting) [gru, right of=Gru1Accounting, xshift=\xShiftModelTikz] {GRU(38)};

\node (InputTimeInvariant) [input,below=1cm of InputAccounting] {Market(10)};

\node (merge) [merge, below right=0.1 and 0.5 cm of Gru2Accounting] {Merge (48)};

\node (dense1) [dense, right of=merge, xshift=\xShiftModelTikz] {D(19)};
\node (dense2) [dense, right of=dense1, xshift=\xShiftModelTikzDense] {D(8)};
\node (dense3) [dense, right of=dense2, xshift=\xShiftModelTikzDense] {D(2)};

\draw [arrow] (InputAccounting) -- (Gru1Accounting);
\draw [arrow] (Gru1Accounting) --(Gru2Accounting);
\draw [arrow] (Gru2Accounting) -| (merge);

\draw [arrow] (InputTimeInvariant) -| (merge);

\draw [arrow] (merge) -- (dense1);
\draw [arrow] (dense1) --   (dense2);
\draw [arrow] (dense2) --(dense3);

\end{tikzpicture}
\caption{Visualization Model Architecture}
\label{paper_2_fig_ModelFlowChart}
\end{figure}

We combine the RNN output with the market data and pass it through the feed-forward layers. The model output consists of two values, a quarterly and a yearly earnings prediction. We add dropout and batch normalization between each layer to reduce the likelihood of overfitting. We zero pad each time dependent accounting input to avoid survivorship biases. To allow for non linear relationships, we add hyperbolic tangent activation functions between the layers. We select 5 years of accounting data to ensure that the model is able to recognize time dependent patterns like seasonal businesses, market cycles or company crises; without harming performance with data which is too old or irrelevant for the actual business. However, we also test other choices for the time series length without changing the results. The same holds for other design choices, like using long short-term memory cells (LSTM, \cite{LSTM}) or different number of hidden neurons. Figure \ref{paper_2_fig_ModelFlowChart} illustrates the model architecture. In the following, we will denote our model with \textit{RNN}.

\subsection{Training}
We use the adaptive moment estimation method of \cite{kingma2014adam}, called Adam, to train the base model with a batch size of 512, a learning rate of 0.0075, exponential decay rates for moment estimation of 0.9 and 0.999 and a dropout rate of 0.001. We evaluate and track the model performance after every epoch on the validation set.  An exponential moving average with smoothing factor 0.2 measures the general trend of the validation score. As soon as the moving average increases from one epoch to another, the training stops. In that way, we prevent the model from overfitting on the training set. We train five models with the same setup, but with different random initial model weights. Our final model prediction is the mean of all trained models, as this design reduces the likelihood of a model that performs above or below average just by chance. The loss function is the mean absolute error. We also test a mean squared error loss, but that choice leads the model to fit unusual, extreme earnings better during training, with the drawback of fitting the common and usual earnings less precisely. As our aim is to train a model which is able to predict market expectations of future earnings, we are interested in estimating ordinary earnings results more precisely at the expense of decreasing the forecasting performance of rather unusual results.

\subsection{Benchmarks}
\label{paper2_sec_benchmarks}
We test our model against three benchmarks. First, we compare our model to quarterly and annual analyst forecasts. For quarterly forecasts, we use the most recent mean analysts forecast prior to the reporting date of the actual quarterly earnings results of a specific company. This approach yields at most four quarterly analyst forecasts per company and year, one for every quarter. Regarding the annual analyst forecasts, we operate the same way. Note that this approach leads to at most four different annual analyst forecasts per year and company, too. All of these annual forecasts refer to the same annual earnings report, but with different levels of information. In that way, we ensure that neither our model nor the analysts have some informational advantage.

The second benchmark is the augmented residual income model of \cite{hess2019incorporating}. We only choose the residual income model, since the results are very similar across all models that were considered in the study. As we have seen in section \ref{paper2_sec_literature_review}, the study of \cite{hess2019incorporating} is the most recent and best performing model within the whole literature stream using cross-section panel regression models, at least if we require no dependency on analyst data. \cite{hess2019incorporating} use all available information up to the most recent, quarterly or annual, earnings result to predict the cumulative quarterly earnings of the coming quarters until the next annual earnings result. Their annual earnings forecast is the sum of the already known quarterly results and the following, predicted quarterly results. We divide the predicted cumulative earnings for future quarters by the number of the future quarters to get a prediction for the next quarterly earnings result. In that way, we also get a benchmark for the quarterly predictions. We know the model of \cite{hess2019incorporating} is not designed for quarterly predictions, but as there is no other literature using cross-sectional models that covers quarterly earnings predictions, we assume this approach to be the best proxy for quarterly earnings predictions of cross-sectional models. We will denote this benchmark with \textit{Regression}.

Lastly, we choose the naive random walk model of \cite{bradshaw2012re} as time-series model benchmark, since the authors find strong results despite its simplicity. The other two mentioned models of \cite{ball2018automated} and \cite{elend2020earnings} would have been possible benchmarks too, but their high data requirements complicate an objective comparison in this study. We denote this benchmark with \textit{Random Walk}.

\subsection{Performance Measures}
\label{paper_2_sec_performance_measures}
We compute all of our performance measures using relative or percentage errors, which means that we divide the difference between true earnings result and model forecast by the true earnings result. The same logic is applied to all benchmarks. The reason for this step is that the difference between the actual and the predicted EPS number alone is not an indication of how much the predicted value has been exceeded. For one company, a difference of one dollar could be very high, while the same difference for another company could be negligible. We assume that the percentage estimation error is more meaningful and enables the comparison of companies with different absolute earnings numbers and different numbers of shares outstanding. 

\subsubsection{Median Percentage Difference \& Median Absolute Percentage Difference}
\label{paper2_sec_mpd_mapd}
Most of the studies mentioned in section \ref{paper2_sec_literature_review} use the forecast bias and accuracy as performance measures for their models. We will use these measures too, but with different names, since bias and accuracy have other meanings in the machine learning literature \citep{sokolova2009systematic}. We apply accuracy and bias with their machine learning interpretation in one of our empirical analyses, which would result in a naming conflict. Therefore, we define the forecast bias as the \textit{Median Percentage Difference (MPD)}. It indicates whether a prediction systematically over- or underestimates a ground truth. Accordingly, we define forecast accuracy as \textit{Median Absolute Percentage Difference (MAPD)}, which measures how precise a model is able to predict future earnings. 

\subsubsection{Sign Prediction}
\label{paper2_sec_sign_pred_def}
The sign prediction tests the models performance in a scenario where we are only interested in the direction of the change of the earnings results within one year. For quarterly results, this approach implies that we compare changes between the same quarters of subsequent years, so that we account for seasonal businesses.  The target is a categorical variable representing either a negative change, no change or a positive change in the true earnings results. We define that no change occurs if the absolute value of the relative earnings change is smaller than 5\%, e.g. $|\frac{EPS_t-EPS_{t-1}}{EPS_{t-1}}|<0.05$. In the same way, we define the categorical earnings change predictions for our model and the benchmarks.

We compute average accuracy, macro precision, macro recall and macro F1 as defined in \cite{sokolova2009systematic} for our model and all benchmarks. We think that the ability of a model to predict the direction of an earnings change is a key feature since the direction of an earnings change is an important proxy of whether a fiscal year was successful or not. 

\subsubsection{Earnings Response Coefficient}
\label{paper2_sec_ERC_def}
According to the efficient market hypothesis of \cite{fama1970}, stock prices should react quickly to surprising earnings announcements. We define an earnings surprise as a situation where the market expectation of an earnings results is different to the realized earnings. Therefore, if the market reacts accordingly to high deviations between model prediction and realized earnings, we conclude that the model prediction is a suitable proxy for the market expectation. To do so, we first estimate the abnormal returns at the earnings announcement dates with a three day event window using the market model:
\begin{equation}
r^i_t-r^f_t=\alpha_0+\alpha_1(r^m_t-r^f_t)+\sum_j \alpha_2^{i,j} D^{i,j}_t+\epsilon^i_t,
\end{equation}
where $r^i$ is the return of company $i$, $r^f$ is the risk-free rate (3 months LIBOR), $r^m$ is the return of the S\&P 500 index and $D^{i,j}$ is the event dummy of company $i$ at earnings announcement $j$. Accordingly, $\alpha_2^{i,j}$ is the cumulative abnormal return of company $i$ and earnings announcement $j$.

Finally, we estimate a year- and firm-fixed effects model with clustered standard errors, specified as follows:
\begin{equation}
\begin{aligned}
\alpha_2^{i,j}= & \beta_0 surprise_{i,j}\\
 & + \beta_1 ln(TotalAssets_{i,j})+ \beta_2 ln(TotalAssets_{i,j}) \times surprise_{i,j}\\
 & + \beta_3 TobinsQ_{i,j}+ \beta_4 TobinsQ_{i,j} \times surprise_{i,j}\\
& + \sum_y D_{i,j}^y + \epsilon_{i,j},
\end{aligned}
\end{equation}
where the earnings surprise is defined as $\frac{EPS_{i,j}-\widehat{EPS}_{i,j}}{\widehat{EPS}_{i,j}}$ with $EPS_{i,j}$ being the realized earnings per share of company $i$ and earnings announcement $j$, whereas $\widehat{EPS}_{i,j}$ is its prediction. Coefficient $\beta_0$ is the so called \textit{Earnings Response Coefficient (ERC)}. The higher the ERC, the stronger is the relationship between the estimated earnings surprise and the market reaction. $D_{i,j}^y$ is the year dummy of year $y$ and announcement $j$ of company $i$. For quarterly results, we also add quarterly dummies. 

\section{Data}
\label{paper2_sec_data}    
\subsection{Data Selection}
\label{paper2_sec_data_selection} 
Our sample includes all NYSE, Amex and Nasdaq listed securities with stock code 10 or 11 that are in the intersection of the CRSP annual returns file and the Compustat fundamentals annual file between 1990 and 2021. The analyst data originates from the I/B/E/S summary file.

In order to be comparable with other studies, we adapt the accounting data selection from the literature stream using cross-sectional models as \cite{fama2000forecasting}, \cite{hou2012implied}, \cite{li2014evaluating}, \cite{hess2019incorporating} or \cite{azevedo2021earnings}. Accordingly, our earnings number is the earnings before extraordinary items, but after dividends paid on preferred stock. As explanatory variable we use lagged earnings as former studies find evidence that this variable has a high explanatory power for future earnings (\cite{bradshaw2012re}). We measure a companies financial stability using the equity ratio, defined as stockholders equity divided by Compustat's total assets. Since \cite{lewellen2019accruals} find a negative relationship between accruals and subsequent earnings, we also add accruals, defined as by \cite{hou2012implied}. We add total cash dividends as input to our model. Finally, we use total assets as a proxy for size. We scale earnings, dividends and accruals by the weighted average common shares outstanding. The accounting input is a time series of the last 20 quarters for which all of the mentioned variables are available. We use zero padding in cases without sufficient data. The minimum requirement is one observation, to ensure the model to be minimal data intensive. Since occasional quarters can be missing, we add a time variable that indicates how many quarters have passed, expressed as a fraction of a year. In that way we ensure that the model is still able to learn seasonal and economic cycles.

Our company specific market variables are the individual stock returns, volume per share and Tobins's q. The former two variables are proxies for the companies general development and the markets interest in the company. Tobin's q is defined as:
\begin{equation}
Tobin's\ q = \frac{Market\:Value\:of\:Equity\:+\:Book\:Value\:of\:Liabilities}{Book\:Value\:of\:Equity\:+\:Book\:Value\:of\:Liabilities}
\end{equation}
It is a proxy for the change in market expectations on future earnings of the company.
In addition, we incorporate market wide variables as the S\&P 500 Index to cover market wide developments and the CBOE Volatility Index to cover market stress. We observe these variables over the last quarter (63 days) prior to the earnings announcement we like to predict.

We compress all stock and market variables as follows: We compute quarterly log returns of the companies' stock price and the market index. We cumulate the companies' volume per share. Tobin's q and the Volatility index are captured as the absolute difference between the start and end of the quarter. If there are less than 63 days of information available, we work with the maximum period length for which data is available. As the period length might have an influence on all the mentioned market variables, we express these market variables on a per-day basis. We add a time variable which denotes the length of the time period. 

Finally, we add quarter dummies to make sure that the model is able to learn seasonal effects.
We log transform total assets, total cash dividends and the absolute changes in the Volatility Index and Tobins'q to reduce the skewness in these variables. We winsorize all variables at the 1\textit{st} and 99\textit{th} percentile to minimize the effect of outliers. Additionally, we studentize all non-binary inputs as:
\begin{equation}
\tilde{x_i}=\frac{x_i-\bar{x}}{\sqrt{\frac{1}{n}\sum_i(x_i-\bar{x})^2}},
\end{equation}
where $\bar{x}$ is the mean of $x_i$.
\subsection{Data Preparation}
\begin{table}
\begin{flushleft}
\caption{Data Preparation}
 \mytablefont{This table presents the data preparation steps for the RNN model and the regression model as defined in section \ref{paper2_sec_methodology}. The upper half of the table consists of cleaning steps that belong to both models. The lower half of the table represents steps that are necessary to the specific models.}
\label{paper2_tab_preprocess_data}
\mytablefont
\begin{tabularx}{\textwidth}{ *{6}{Y}}
\toprule
\multicolumn{2}{c}{\textbf{Preparation Step}} & \multicolumn{2}{c}{\textbf{Num. Obs.}} & \multicolumn{2}{c}{\textbf{Num. Obs. Removed}} \\
\midrule
\multicolumn{2}{c}{Raw Compustat Data} & \multicolumn{2}{c}{1,826,340} & \multicolumn{2}{c}{} \\
\multicolumn{2}{c}{No CRSP Match} & \multicolumn{2}{c}{899,662} & \multicolumn{2}{c}{-926,678} \\
\multicolumn{2}{c}{Missing Year Or Quarter} & \multicolumn{2}{c}{899,379} & \multicolumn{2}{c}{-283} \\
\multicolumn{2}{c}{\parbox{5cm}{\centering Missing Total Assets Or Weighted Average Common Shares}} & \multicolumn{2}{c}{736,884} & \multicolumn{2}{c}{-162,495} \\
\multicolumn{2}{c}{Missing EPS Or BVE} & \multicolumn{2}{c}{706,601} & \multicolumn{2}{c}{-30,283} \\
\midrule
\multicolumn{3}{c}{\textbf{RNN}} & \multicolumn{3}{c}{\textbf{Regression}} \\
Step & Num. Obs. & Num. Obs. Removed & Step & Num. Obs. & Num. Obs. Removed \\
\cmidrule(lr){1-3}  \cmidrule(lr){4-6} 
Remove Yearly Data & 563,654 & -142,947 & Remove Incomplete Years & 695,640 & -10,961 \\
Remove Duplicate Quarters & 563,427 & -227 & Remove Q4 & 556,512 & -139,128 \\
Require >1 Obs. Per Comp. & 563,291 & -136 & Remove Year Gaps & 496,320 & -60,192 \\
Remove Comp. First Obs. & 548,679 & -14,612 &  &  &  \\
Remove NaN Or Inf. Obs. & 419,432 & -129,247 &  &  &  \\
\bottomrule
\end{tabularx}
\end{flushleft}
\end{table}

Table \ref{paper2_tab_preprocess_data} lists all data preparation steps. The upper half of the table describes general steps that we apply to the raw Compustat data. We start with 1,826,340 earnings observations, which contain both annual and quarterly earnings results. After conducting essential cleaning steps, as coverage in CRSP, filtering missing information about the year or quarter of the observation or filtering observations with missing total assets, shares outstanding, earnings or book value of equity, we end up with 706,601 observations.

As firm coverage is a key point in our analysis, we emphasize the data intensity of our model and compare it with the benchmark regression model. Therefore, the lower half of Table \ref{paper2_tab_preprocess_data} compares all cleaning steps for both models separately. Regarding the RNN, we first remove annual observations as we only consider quarterly results for our earnings forecasts.\footnote{We employ historical quarterly earnings data to forecast both future quarterly and annual earnings outcomes. To ensure accurate annual predictions based on quarterly results, we make the fundamental assumption that the summation of quarterly diluted earnings per share is equivalent to the annual diluted earnings per share. Notably, our findings reveal that the absolute limits of the 90\% confidence interval for the disparity between annual EPS and the sum of quarterly EPS are below 0.2.} We minimize the data requirements by allowing inputs with only one historical observation. This implies that we remove companies that only have one observation in the sample. We also remove the first observation of each company since these observations have no predecessors which can serve as predictors. Finally, we remove duplicates and missing observations.

Regarding the cross-sectional approach of \cite{hess2019incorporating}, we have to remove all Q4 observations by design of their model. This step is comparable to our removal of annual observations. As the authors use all the earnings data of one year to predict the next year, they have to remove incomplete or missing years out of their sample.

We end up with a sample size of 419,432 for our model compared to a sample size of 496,320 for the benchmark model, which indicates that the data intensity of our model is comparable to the one of the benchmark. This is in contrast to prior time-series models, which have in general higher data requirements than cross-sectional models. Table \ref{paper2_tab_sumstat} presents the summary statistics of the accounting and market data.

\begin{table}
\begin{flushleft}
\caption{Summary Statistics}
 \mytablefont{This table presents the summary statistics of the accounting data (upper half) and market data (lower half) as defined in section \ref{paper2_sec_data_selection}. We compute selected percentiles as well as mean, standard deviation and number of observations. The accounting data is given to our model as time-series of the length of 20, which is why we get more than 419,432 observations, as stated in Table \ref{paper2_tab_preprocess_data}. The market variables contain compressed information about the last quarter before earnings announcement, defined as follows: We cumulate daily stock returns, market index returns and daily volume. We calculate the absolute change of volatility index and Tobin's q. The period length, expressed as a fraction of a year, is the minimum of 63 days and the number of days for which information is available. Since some periods are shorter than 63 days, we normalize all market variables (except the period length) on a daily basis. We report the market variables as percentage numbers, except for the period length.}
\label{paper2_tab_sumstat}
\mytablefont
\begin{tabularx}{\textwidth}{L{3.5cm}c *{8}{Y}}
\toprule
 & \textbf{Count} & \textbf{Mean} & \textbf{Std} & \textbf{1\%} & \textbf{25\%} & \textbf{50\%} & \textbf{75\%} & \textbf{99\%} \\
\midrule
  \multicolumn{9}{c}{\textbf{Accounting Data}} \\
       \midrule
Earnings Per Share & 443,023 & 0.01 & 0.96 & -4.93 & -0.10 & 0.07 & 0.27 & 2.87 \\
Total Assets (Billions) & 443,023 & 2.78 & 13.80 & 0.00 & 0.05 & 0.20 & 1.03 & 48.01 \\
Equity Ratio & 443,023 & 0.51 & 0.26 & -0.18 & 0.34 & 0.52 & 0.71 & 0.94 \\
Dividend Per Share & 443,023 & 0.13 & 0.34 & 0.00 & 0.00 & 0.00 & 0.07 & 1.80 \\
Accruals Per Share & 443,023 & -0.55 & 1.70 & -8.54 & -0.69 & -0.13 & 0.05 & 3.16 \\
Time To Announcement & 6,757,159 & 2.41 & 1.44 & 0.21 & 1.17 & 2.26 & 3.58 & 5.02 \\
\midrule
  \multicolumn{9}{c}{\textbf{Market Data}} \\
       \midrule
Average Daily Stock Return (\%) & 419,432 & 0.09 & 0.47 & -1.23 & -0.15 & 0.07 & 0.30 & 1.76 \\
Average Daily Volume Per Share (\%) & 419,432 & 0.74 & 0.85 & 0.02 & 0.21 & 0.47 & 0.93 & 5.21 \\
Average Daily Market Return (\%) & 419,432 & 0.04 & 0.12 & -0.33 & -0.00 & 0.06 & 0.12 & 0.26 \\
Average Daily Change Volatility Index (\%) & 419,432 & -0.01 & 10.93 & -35.00 & -4.38 & -0.59 & 3.38 & 36.95 \\
Average Daily Change Tobin's q (\%) & 419,432 & -0.12 & 2.48 & -14.50 & -0.35 & -0.00 & 0.33 & 10.30 \\
Period Length & 419,432 & 0.25 & 0.02 & 0.23 & 0.25 & 0.25 & 0.26 & 0.27 \\
\bottomrule
\end{tabularx}
\end{flushleft}
\end{table}

We split the data set in a training set, a test set and a validation set. The training set contains the oldest 70\% of the earnings results, based on the report date, which results in 293,605 observations between 1990 and 2010. This is the data on which we train our model to adjust the models' weights. The test set contains the latest 20\% of all earnings results, which implies 83,732 observations between 2014 and 2021. We use the test set to evaluate the performance of the final model and all benchmarks. Accordingly, the validation set contains the remaining 10\% of the data, so that this set contains 42,095 observations between 2010 and 2014. The purpose of the validation set is to find the best hyperparameter setup for our model. The reason why we do not use a random split of the data is that through the use of market wide predictors like the S\&P 500 or the volatility index, the model might be able to learn patterns in market wide events like the Covid pandemic and may apply this knowledge to out-of-sample earnings results within the time frame of the market wide event.

\FloatBarrier
\section{Empirical Results}
\label{paper2_sec_results}
This section summarizes our study's empirical results. Sections \ref{paper2_sec_overall_performance} and \ref{paper2_sec_quarterly_performance} compare the model with all benchmarks with respect to the median absolute percentage difference and the median percentage difference. The former section yields the results for the whole test set, the latter splits them on a quarterly basis. Section \ref{paper2_sec_firm_size} lists the results per firm sizes, measures by total assets, whereas Section \ref{paper2_sec_industry} compares the performance between different industries. Section \ref{paper2_sec_sign_prediction} covers the sign prediction performance. Sections \ref{paper2_sec_covered_uncovered} and \ref{paper2_sec_covid} each compare the models and benchmarks performance in two contrastive scenarios: Section \ref{paper2_sec_covered_uncovered} compares the performance results for firms with analyst coverage with firms without analyst coverage. Section \ref{paper2_sec_covid} contrasts the performance measures before the Covid crisis with the measures during the crisis. Finally, section \ref{paper2_sec_ERC} contains the results for the earnings response coefficient of all models.

To make the models as comparable as possible, we filter all observations where some data is missing for one or more models. Furthermore, we remove penny stocks, which are observations for which the respective company has a stock price smaller than 5 \$. Finally, as we use percentage errors, we reduce the test set by the 5\% smallest and largest prediction errors, since in cases where the predicted earnings result is very small or even zero, the percentage error gets extraordinary high. 
\subsection{Overall Performance}
\label{paper2_sec_overall_performance}
\begin{table}
\begin{flushleft}
\caption{Overall Performance}
 \mytablefont{This table presents the median absolute percentage difference (MAPD) and the median percentage difference (MPD) between realised and predicted earnings per share, as defined in section \ref{paper2_sec_mpd_mapd}, for the RNN model (section \ref{paper2_sec_model}) and all benchmarks. The benchmarks are analyst forecasts, the cross-sectional regression model of \cite{hess2019incorporating} and the random walk model \citep{bradshaw2012re}, see section \ref{paper2_sec_benchmarks}. We report the results for annual and quarterly predictions. We test for every benchmark whether its median value is significantly different to the median value of the RNN model using the Wilcoxon rank test \citep{wilcoxon1992individual}.}
\label{paper2_tab_overall_performance}
\mytablefont
\begin{tabularx}{\textwidth}{m{3cm} *{4}{Y}}
\toprule
 & \multicolumn{2}{c}{\textbf{Annual}} & \multicolumn{2}{c}{\textbf{Quarterly}} \\
 & \scriptsize{\shortstack{Median Absolute \\ Percentage Difference}} & \scriptsize{\shortstack{Median Percentage \\ Difference}} & \scriptsize{\shortstack{Median Absolute \\ Percentage Difference}} & \scriptsize{\shortstack{Median Percentage \\ Difference}} \\
\midrule
RNN & 19.39 \% & 1.03 \% & 30.74 \% & 1.38 \% \\
Analyst & 25.49 \% & -13.47 \% & 26.67 \% & -7.10 \% \\
Regression & 28.16 \% & 5.86 \% & 50.33 \% & 11.95 \% \\
Random Walk & 34.43 \% & 5.94 \% & 37.55 \% & 2.42 \% \\
Wilcoxon Rank P-Val. RNN-Analyst & 0.00 \% & 0.00 \% & 0.00 \% & 0.00 \% \\
Wilcoxon Rank P-Val. RNN-Regression & 0.00 \% & 0.00 \% & 0.00 \% & 0.00 \% \\
Wilcoxon Rank P-Val. RNN-RW & 0.00 \% & 0.00 \% & 0.00 \% & 0.00 \% \\
Num. Obs. & \multicolumn{2}{c}{50,573} & \multicolumn{2}{c}{50,324} \\
\bottomrule
\end{tabularx}
\end{flushleft}
\end{table}

Table \ref{paper2_tab_overall_performance} contains the overall performance with respect to MAPD and MPD (see section \ref{paper2_sec_mpd_mapd}) of our model and the benchmarks on the test set. Regarding the median absolute percentage error, we see that our model outperforms all benchmarks by at least 6 percentage points for the annual predictions. Additionally, our model is with a median percentage difference of 1.03\% less biased than the analyst forecasts with -13.47\% and the other benchmarks with 5.86\% and 5.94\%, respectively. Note that the negative bias in analyst forecasts implies an 
overly optimistic expectation of future earnings, which is in line with the prior literature \cite[see][]{abarbanell1991analysts,o1988analysts,bradshaw2012re}. 

For quarterly predictions, we see that the RNN model is less precise than the analysts. However, both the random walk model and the cross-sectional regression are outperformed by more than 6 and 19 percentage points, respectively. Analyst forecasts are, with a median percentage difference of -7.10\%, more biased than our model with 1.38\%. The other models are also more biased. To test whether the mentioned differences are statistically significant, we perform Wilcoxon signed-rank tests \citep{wilcoxon1992individual}. Wee see that all differences are statistically significant at the 1\% level. 
\subsection{Performance Per Quarter}
\label{paper2_sec_quarterly_performance}
\begin{table}
\begin{flushleft}
\caption{Performance Per Quarter}
 \mytablefont{This table presents the median absolute percentage difference (MAPD) and the median percentage difference (MPD) between realised and predicted earnings per share, as defined in section \ref{paper2_sec_mpd_mapd}, partitioned by fiscal quarter. We compare the RNN model (section \ref{paper2_sec_model}) with analysts forecasts, the cross-sectional regression model of \cite{hess2019incorporating} and the random walk model \citep{bradshaw2012re}, see section \ref{paper2_sec_benchmarks}. We report the results for annual and quarterly predictions. For annual predictions, we test for all models whether their median value in quarter 4 is significantly different to the median value of quarter 1 using the Mann-Whitney U test \citep{mann1947test}.}
\label{paper2_tab_quarterly_performance}
\mytablefont
\begin{tabularx}{\textwidth}{c *{10}{Y}}
\toprule
 & \multicolumn{2}{r}{\textbf{Quarter 1}} & \multicolumn{2}{r}{\textbf{Quarter 2}} & \multicolumn{2}{r}{\textbf{Quarter 3}} & \multicolumn{2}{r}{\textbf{Quarter 4}} & \multicolumn{2}{r}{\textbf{P Val. 1-4}} \\
 & MAPD & MPD & MAPD & MPD & MAPD & MPD & MAPD & MPD & MAPD & MPD \\
\midrule
   \multicolumn{11}{c}{\textbf{Panel A: Annual}} \\
       \midrule
RNN (\%) & 32.5 & 2.0 & 23.0 & 1.2 & 16.4 & 0.9 & 11.5 & 0.8 & 0.0 & 15.9 \\
Analyst (\%) & 30.6 & -16.4 & 27.4 & -14.5 & 24.0 & -12.9 & 20.5 & -11.0 & 0.0 & 7.4 \\
Regression (\%) & 39.9 & 4.4 & 33.7 & 6.2 & 26.6 & 6.9 & 18.0 & 5.7 & 0.0 & 8.0 \\
Random Walk (\%) & 34.2 & 5.1 & 34.0 & 5.7 & 34.4 & 6.3 & 35.0 & 6.5 & 1.1 & 3.3 \\
Num. Obs. & \multicolumn{2}{c}{11,503} & \multicolumn{2}{c}{12,780} & \multicolumn{2}{c}{13,022} & \multicolumn{2}{c}{13,268} & &  \\
\midrule
   \multicolumn{11}{c}{\textbf{Panel B: Quarterly}} \\
       \midrule
RNN (\%) & 33.7 & -1.4 & 27.9 & 0.0 & 26.6 & 2.8 & 36.2 & 3.6 &  &  \\
Analyst (\%) & 25.9 & -6.5 & 25.4 & -6.7 & 23.7 & -5.6 & 32.8 & -10.6 &  &  \\
Regression (\%) & 42.5 & -7.0 & 43.7 & 9.2 & 49.5 & 20.5 & 61.7 & 31.5 &  &  \\
Random Walk (\%) & 46.8 & -4.4 & 34.3 & 10.1 & 30.6 & 3.9 & 41.8 & -1.8 &  &  \\
Num. Obs. & \multicolumn{2}{c}{11,368} & \multicolumn{2}{c}{12,873} & \multicolumn{2}{c}{13,063} & \multicolumn{2}{c}{13,020} & &  \\
\bottomrule
\end{tabularx}
\end{flushleft}
\end{table}

Table \ref{paper2_tab_quarterly_performance} presents the overall performance results, separated per quarter. For annual predictions, we recognize that the median absolute percentage errors of the RNN, the regression model and the analysts forecasts decrease with increasing quarter. This seems intuitive since the amount of information increases during a fiscal year, leading to more precise predictions. In the same vein, it is plausible that the random walk models is not improving its performance with increasing quarters, as this model makes no use of additional information. However, we see that the analysts performance improvement is not as large as our models improvement with increasing quarters. Between quarter one and four, our model improves its performance about 21 percentage points, whereas the analysts only improve about 10 percentage points. This indicates that our model is more efficient in processing new information during a fiscal year than the analysts. Looking at the bias of the annual forecasts, we see an improvement with increasing quarters too, from 2\% in the first quarter to 0.8\% in the fourth quarter. The analyst forecasts also show less bias in the later quarters, although the absolute level of bias remains strictly higher in all quarters compared to the RNN forecasts. The other benchmark models do not decrease their forecast bias at all. To test if there is a statistically significant improvement between quarter one and four, we conduct a Mann-Whitney U test \citep{mann1947test}. For the accuracy of the predictions, we see that the differences between the first and the last quarter of all but the random walk model are statistically significant at the 1\% level. Regarding the bias of the predictions, there is no statistically significant difference for the RNN model, which is likely due to the already low bias in the first quarter.

Looking at the quarterly predictions, we observe for all but the regression model the general trend that quarters one and four seem to be harder to predict than quarters two and three, since the median absolute percentage errors are higher for quarter one and four. We also observe that the weaker performance on quarterly results of our model compared to analysts stems mostly from the first quarter. Although the precision of RNN predictions is slightly lower than that of the analyst forecasts in quarters two, three and four, the results demonstrate a substantial gap of 7.8 percentage points during the first quarter. Looking at the median percentage difference, we see that our model is in all quarters less biased than all other benchmarks, with the exception of the random walk in quarter four.

For the following analyses, we move the quarterly results to the appendix, since similar patterns emerge as on an annual basis.
\subsection{Performance Per Firm Size}
\label{paper2_sec_firm_size}
\begin{table}
\begin{flushleft}
\caption{Annual Performance on Different Firm Sizes (Median Absolute Percentage Difference)}
 \mytablefont{This table presents the median absolute percentage difference (MAPD) between realised and predicted annual earnings per share, as defined in section \ref{paper2_sec_mpd_mapd}, partitioned by firm size. We define size deciles using total assets. We compare the RNN model (section \ref{paper2_sec_model}) with analyst forecasts, the cross-sectional regression model of \cite{hess2019incorporating} and the random walk model \citep{bradshaw2012re}, see section \ref{paper2_sec_benchmarks}.}
\label{paper2_tab_firm_size_prediction_accuracyY}
\mytablefont
\begin{tabularx}{\textwidth}{c *{6}{Y}}
\toprule
\textbf{Size Decile} & \textbf{RNN} & \textbf{Analyst} & \textbf{Regression} & \textbf{Random Walk} & \textbf{Num. Obs.} \\
\midrule
1 & 19.61 \% & 22.49 \% & 29.81 \% & 36.57 \% & 5,058 \\
2 & 21.30 \% & 30.16 \% & 29.34 \% & 39.55 \% & 5,057 \\
3 & 22.40 \% & 33.56 \% & 30.16 \% & 40.34 \% & 5,057 \\
4 & 21.71 \% & 35.73 \% & 29.50 \% & 38.14 \% & 5,057 \\
5 & 20.37 \% & 29.54 \% & 27.95 \% & 35.20 \% & 5,057 \\
6 & 20.82 \% & 28.69 \% & 28.65 \% & 35.88 \% & 5,058 \\
7 & 20.15 \% & 24.96 \% & 27.95 \% & 33.81 \% & 5,057 \\
8 & 18.01 \% & 21.94 \% & 26.48 \% & 31.02 \% & 5,057 \\
9 & 16.14 \% & 20.27 \% & 26.16 \% & 28.06 \% & 5,057 \\
10 & 13.91 \% & 13.81 \% & 25.95 \% & 25.14 \% & 5,058 \\
\bottomrule
\end{tabularx}
\end{flushleft}
\end{table}

Next, we investigate the performance of all models with respect to different firm sizes, measured by total assets. To do so, we split the test set in size deciles and we compute the median absolute percentage error per decile and model. Table \ref{paper2_tab_firm_size_prediction_accuracyY} shows the results. As a general pattern, the models tend to exhibit a decline in performance in the first three to four deciles, depending on the specific model. However, performance subsequently improves in the following deciles. Our model outperforms all benchmarks for all deciles, with the only exception being the analyst forecasts in the last decile (13.91\% vs. 13.81\%). It should be noted that the performance difference between the RNN model and the analyst forecasts ranges between 8 and 14 percentage points for deciles 2-6, indicating that our model surpasses the analyst forecasts particularly for small to medium-sized companies.

\subsection{Performance Per Industry}
\label{paper2_sec_industry}
\begin{table}
\begin{flushleft}
\caption{Annual Performance on Different Industries (Median Absolute Percentage Difference)}
 \mytablefont{This table presents the median absolute percentage difference (MAPD) between realised and predicted annual earnings per share, as defined in section \ref{paper2_sec_mpd_mapd}, partitioned by industry. We use the economic sector class of The Refinitiv Business Classifications (TRBC) for industry classification. We compare the RNN model (section \ref{paper2_sec_model}) with analyst forecasts, the cross-sectional regression model of \cite{hess2019incorporating} and the random walk model \citep{bradshaw2012re}, see section \ref{paper2_sec_benchmarks}.}
\label{paper2_tab_industry_prediction_accuracyY}
\mytablefont
\begin{tabularx}{\textwidth}{c *{6}{Y}}
\toprule
\textbf{Industry} & \textbf{RNN} & \textbf{Analyst} & \textbf{Regression} & \textbf{Random Walk} & \textbf{Num. Obs.} \\
\midrule
Academic \& Educational Services & 16.35 \% & 24.77 \% & 29.74 \% & 29.68 \% & 238 \\
Basic Materials & 24.93 \% & 26.78 \% & 32.88 \% & 44.54 \% & 2,999 \\
Consumer Cyclicals & 20.20 \% & 21.63 \% & 27.67 \% & 34.02 \% & 9,319 \\
Consumer Non-Cyclicals & 15.22 \% & 19.14 \% & 22.05 \% & 25.67 \% & 2,946 \\
Energy & 47.93 \% & 50.61 \% & 51.24 \% & 87.90 \% & 2,458 \\
Financials & 15.60 \% & 20.24 \% & 24.80 \% & 26.12 \% & 740 \\
Healthcare & 16.77 \% & 22.19 \% & 27.90 \% & 31.66 \% & 10,335 \\
Industrials & 17.46 \% & 19.45 \% & 25.69 \% & 29.68 \% & 8,889 \\
Other & 29.95 \% & 22.70 \% & 27.50 \% & 51.29 \% & 40 \\
Real Estate & 23.43 \% & 22.95 \% & 30.44 \% & 41.11 \% & 329 \\
Technology & 21.83 \% & 49.69 \% & 31.19 \% & 41.62 \% & 10,350 \\
Utilities & 10.74 \% & 7.38 \% & 20.78 \% & 15.03 \% & 1,930 \\
\bottomrule
\end{tabularx}
\end{flushleft}
\end{table}

Table \ref{paper2_tab_industry_prediction_accuracyY} contains the MAPD results for different industries. For industry classification, we use the economic sector class of The Refinitiv Business Classifications (TRBC). If there is no class label for a given observation, we assign it to the "Other" class. The results of the previous analyses confirm in a way that the RNN model outperforms all benchmarks for nine out of twelve industries. The three exceptions are the real estate and the utilities sector as well as the others category. For the first two of these exceptions, the differences are all smaller than 3.5 percentage points. For the others class, the difference is higher, but that class is very small with only 40 observations. It is noticeable that the median absolute percentage errors in the energy sector are clearly higher compared to all other industries (except "Other"). This effect holds for all models.

\subsection{Sign Prediction}
\label{paper2_sec_sign_prediction}
\begin{table}
\begin{flushleft}
\caption{Sign Prediction (Annual)}
 \mytablefont{This table presents the results for the sign prediction analysis for annual data, as defined in section \ref{paper2_sec_sign_pred_def}, for the RNN model (section \ref{paper2_sec_model}) and all benchmarks. This analysis examines the models ability to predict the direction of an earnings change from the previous to the actual period. We define three classes: positive (percentage change > 5\%), negative (percentage change < -5\%) and neutral (|percentage change| < 5\%). The benchmarks are analysts forecasts, the cross-sectional regression model of \cite{hess2019incorporating} and the random walk model \citep{bradshaw2012re}, see section \ref{paper2_sec_benchmarks}. We calculate macro accuracy, precision, recall and F1 for all models, as defined in \cite{sokolova2009systematic}.}
\label{paper2_tab_sign_predictionY}
\mytablefont
\begin{tabularx}{\textwidth}{c *{5}{Y}}
\toprule
 & \textbf{RNN} & \textbf{Analyst} & \textbf{Random Walk} & \textbf{Regression} \\
\midrule
Accuracy & 80.57 \% & 75.81 \% & 40.93 \% & 72.91 \% \\
Precision & 63.60 \% & 52.97 \% & 11.40 \% & 49.53 \% \\
Recall & 65.50 \% & 52.61 \% & 33.33 \% & 48.74 \% \\
F1 & 64.06 \% & 52.72 \% & 20.47 \% & 47.55 \% \\
Num. Obs. & \multicolumn{4}{c}{42,525} \\
\bottomrule
\end{tabularx}
\end{flushleft}
\end{table}

Table \ref{paper2_tab_sign_predictionY} shows the results of the sign prediction analysis, as explained in section \ref{paper2_sec_sign_pred_def}. We see that our model outperforms all other benchmarks with respect to accuracy, precision, recall and F1. While the accuracy is almost 5 percentage points better than that of the analysts, our model even outperforms the analysts by more than 10 percentage points regarding the other three measures. The difference becomes even clearer compared to the other two benchmarks.

\subsection{Performance Covered vs. Uncovered Firms}
\begin{table}
\begin{flushleft}
\caption{Performance Analyst Coverage vs. Similar No Analyst Coverage (Annual)}
 \mytablefont{This table presents the median absolute percentage difference (MAPD), as defined in section \ref{paper2_sec_mpd_mapd}, of the annual RNN model EPS predictions (see section \ref{paper2_sec_model}) and realised EPS values for not analyst-covered firms and similar analyst-covered firms. Two firms are similar if they are in the same industry and if the eucledian distance between their arrays consisting of standarized values of total assets and Tobin's q is smaller than 0.01. The sample consists of all non-covered firm observations for which there is at least one similar covered firm observation. If there is more than one similar firm observation, we take the most similar one. We test whether the median value for non-covered firms is significantly different to the median value for similar covered firms using the Wilcoxon rank test \citep{wilcoxon1992individual}.}
\label{paper2_tab_covered_similar_uncoveredY}
\mytablefont
\begin{tabularx}{\textwidth}{m{2cm} *{4}{Y}}
\toprule
 & Count & \shortstack{MAPD \\ Covered} & \shortstack{MAPD \\ Uncovered} & P Value \\
\midrule
Academic \& Educational Services & 78 & 20.3 \% & 28.1 \% & 79.4 \% \\
Basic Materials & 809 & 27.0 \% & 25.7 \% & 7.9 \% \\
Consumer Cyclicals & 4,189 & 25.2 \% & 33.9 \% & 0.0 \% \\
Consumer Non-Cyclicals & 520 & 28.7 \% & 19.4 \% & 10.7 \% \\
Energy & 846 & 53.9 \% & 49.7 \% & 90.3 \% \\
Financials & 135 & 31.0 \% & 33.3 \% & 68.5 \% \\
Healthcare & 6,798 & 16.2 \% & 16.6 \% & 0.1 \% \\
Industrials & 4,201 & 22.7 \% & 22.9 \% & 22.6 \% \\
Other & 6 & 73.2 \% & 116.7 \% & 15.6 \% \\
Real Estate & 46 & 23.5 \% & 42.4 \% & 87.1 \% \\
Technology & 4,878 & 27.7 \% & 32.3 \% & 0.0 \% \\
Utilities & 159 & 8.7 \% & 11.9 \% & 0.6 \% \\
 \midrule
Total & 22,665 & 22.3 \% & 24.8 \% & 0.0 \% \\
\bottomrule
\end{tabularx}
\end{flushleft}
\end{table}

\label{paper2_sec_covered_uncovered}
We have already seen that our model performs better than analyst forecasts if we consider annual earnings forecasts. However, all of the previous results are evaluated on a sample for which observations for every model are available. A natural question is whether our model performs equally well for observations without analyst coverage. As a first approach, we investigate the difference in performance of our model and the remaining benchmarks in both cases, analyst coverage and no analyst coverage. Table \ref{paper2_tab_covered_uncovered} in the appendix shows the results. For annual predictions, we see that the MAPD of our model increases about 9 percentage points if we consider data that is not covered by analysts. A similar decrease in performance can be observed for the benchmarks, too. Regarding the MPD, we observe the opposite pattern. For all models, the predictions get less biased if the data is not covered by analysts. All of these effects are statistically significant on the 5\% level.

A drawback of the approach of Table \ref{paper2_tab_covered_uncovered} is that both samples, i.e. the covered and uncovered firms, are structurally different. As \cite{li2014evaluating} argue in their study and which is in line with our findings, analyst forecasts are available only for a subset of firms, with almost half of all firms not having analyst coverage in most years. Most of the firms without analyst coverage are typically small and young firms. Therefore, in a second step we try to reduce the structural difference between covered and uncovered firms, so that we end up with a clean comparison of the performance of our model in both situations. The idea is to only compare covered with uncovered firms that are similar with respect to industry, total assets and Tobin's q. To achieve this, both companies must belong to the same industry. In addition, the Euclidean distance between the vectors consisting of the standardized values of total assets and Tobin's q must be smaller than a certain limit, which we set to 0.01. Within all the uncovered companies that fulfill that property, we take the one with the minimal distance. If there is no uncovered company that fulfills this property, the covered observation will be removed from the analysis. 

Table \ref{paper2_tab_covered_similar_uncoveredY} shows the results of that analysis. To keep things well arranged, we only report the MAPD results for our RNN model. We see that for annual predictions the increase in the median absolute percentage difference is now only 2.5 percentage points instead of 9 percentage points as in Table \ref{paper2_tab_covered_uncovered}.  Even if the differences between covered and uncovered firms are still statistically significant at the 1\% level, we see that considering similar firms increases the models performance on uncovered data strongly. Looking at the specific industries, we see that only for four out of twelve industries, the RNN performs statistically significantly worse (at the 10\% level) if a firm is not covered by analysts. For all other industries, the difference is either insignificant, or, in case of the industry \textit{Basic Materials}, the predictive performance of the RNN is even statistically significantly better for uncovered firms. 
This indicates that our model yields robust results even for firms not covered by analysts. 

\subsection{Performance Pre vs. During Covid}
\label{paper2_sec_covid}
\begin{table}
\begin{flushleft}
\caption{Performance Pre vs. During COVID 19 (Annual)}
 \mytablefont{This table compares the median absolute percentage difference (MAPD) and the median percentage difference (MPD), as defined in section \ref{paper2_sec_mpd_mapd}, for annual earnings reports before and during the Covid pandemic. We compute these measures for the RNN model (section \ref{paper2_sec_model}), analyst forecasts, the cross-sectional regression model of \cite{hess2019incorporating} and the random walk model \citep{bradshaw2012re}, see section \ref{paper2_sec_benchmarks}. We set February 18, 2020 as the starting point for the pandemic. For all models, we test whether their median value for observations prior to the pandemic is significantly different to their median value for observations during the pandemic using the Mann-Whitney U test \citep{mann1947test}.}
\label{paper2_tab_covid_predictionY}
\mytablefont
\begin{tabularx}{\textwidth}{c *{7}{Y}}
\toprule
 & \multicolumn{2}{c}{\textbf{Pre-Covid}} & \multicolumn{2}{c}{\textbf{During-Covid}} & \multicolumn{2}{c}{\textbf{P Value}} \\
 & MAPD & MPD & MAPD & MPD & MAPD & MPD \\
\midrule
RNN & 19.01 \% & 0.78 \% & 20.55 \% & 1.65 \% & 3.69 \% & 0.00 \% \\
Analyst & 24.86 \% & -14.03 \% & 27.25 \% & -11.91 \% & 0.00 \% & 0.00 \% \\
Regression & 27.87 \% & 7.34 \% & 28.89 \% & 1.22 \% & 3.32 \% & 0.00 \% \\
Random Walk & 31.58 \% & 5.74 \% & 41.53 \% & 6.84 \% & 0.00 \% & 0.00 \% \\
Num. Obs. & \multicolumn{2}{c}{37,069} & \multicolumn{2}{c}{13,504} & &  \\
\bottomrule
\end{tabularx}
\end{flushleft}
\end{table}

Another question we would like to answer is how robust our model predictions are during a crisis. The only crisis that occurred during our test data period is the Covid pandemic. Even though other events as the dot-com bubble in 2001, the subprime crisis in 2007/2008 or the European debt crisis in 2009/2010 are contained in our data, the fact that these events are already seen by our model during training would distort the comparison to the benchmark models. We divide the test set in a time prior the Covid pandemic and in a time during the Covid pandemic. We do not include a period that covers the time after the Covid pandemic, since it is not clear whether the pandemic was already over until the end of 2021. We set the 18th of February 2020 as the starting point for the pandemic. 

Looking at Table \ref{paper2_tab_covid_predictionY}, we observe a performance drop with respect to MAPD of 1.5 percentage points for the annual predictions of our model when we focus on observations during the pandemic, indicating that the RNN predictions are robust in times of a crises. The drop in performance is smaller compared to the analysts (2.5 percentage points) and the random walk (10 percentage points) and comparable to the regression model (1 percentage point). Furthermore, our model is slightly more biased during the Covid crisis (1.65\%) than before (0.78\%), but the absolute level is still small. However, the MPD of the other two analysts and the regression model are even decreasing about 2 and 6 percentage points, respectively. We test the performance differences of the models before and during Covid for statistical significance. All differences are statistically significant at the 5\% level.

\subsection{Earnings Response Coefficient}
\label{paper2_sec_ERC}
\begin{table}
\begin{flushleft}
\caption{Earnings Response Coefficient (Annual)}
 \mytablefont{This table presents the annual results for the earnings response regression of section \ref{paper2_sec_ERC}. We fit a year- and firm-fixed effects model with clustered standard errors. Earnings surprise is defined as $\frac{EPS-\widehat{EPS}}{\widehat{EPS}}$ with $EPS$ being the realized earnings per share and $\widehat{EPS}$ is its prediction.}
\label{paper2_tab_ercY}
\mytablefont
\begin{tabularx}{\textwidth}{c *{5}{Y}}
\toprule
 & \textbf{RNN} & \textbf{Analyst} & \textbf{Regression} & \textbf{Random Walk} \\
\midrule
surprise (\%) & 6.15*** & 1.96** & 5.39*** & 0.40 \\
lnTA (\%) & -1.87*** & -1.87*** & -1.89*** & -1.84*** \\
lnAssets x surprise (\%) & -0.56*** & -0.18** & -0.49*** & -0.03 \\
Tobin's q (\%) & -0.31*** & -0.32*** & -0.33*** & -0.31*** \\
Tobin's q x surprise (\%) & -0.01 & -0.04 & -0.05 & -0.03 \\
2015 (\%) & 0.68** & 0.61* & 0.65* & 0.57* \\
2016 (\%) & 1.28*** & 1.21*** & 1.25*** & 1.14*** \\
2017 (\%) & 0.54 & 0.47 & 0.51 & 0.42 \\
2018 (\%) & 1.05** & 1.02** & 1.04** & 0.96** \\
2019 (\%) & 0.93** & 0.89* & 0.99** & 0.83* \\
2020 (\%) & -0.71 & -0.78 & -0.68 & -0.87 \\
2021 (\%) & 0.76 & 0.67 & 0.63 & 0.60 \\
2022 (\%) & 2.28*** & 2.24*** & 2.39*** & 2.17*** \\
$R^2$ (\%) & 2.44 & 1.60 & 2.23 & 1.45 \\
Num. Obs. & 13,180 & 13,180 & 13,180 & 13,180 \\
\bottomrule
\end{tabularx}
\end{flushleft}
\end{table}

 Our last analysis covers the earnings response coefficient, following the approach from section \ref{paper2_sec_ERC_def}. In contrast to the previous analyses, we only consider the most recent prediction prior to the report date of an earnings report. Otherwise, it would not be clear why to expect an announcement effect at the end of a fiscal year if the proxy of market expectation was measured after, for example, quarter one. This step roughly quarters the sample for annual predictions. Another necessary filtering step is the requirement of the existence of the exact report date of an earnings result, as we try to link earnings surprises to stock market effects. Therefore, we remove all observations where the respective Compustat variable (rdq) is not available.
 
Table \ref{paper2_tab_ercY} covers the results of the earnings response analysis. For annual predictions, we see that the earnings response coefficient (ERC), i.e. the coefficient of the earnings surprise variable, is with 6.15\% more than 4 percentage points higher than the ERC of the analysts, almost 1 percentage point higher than the ERC of the regression model and more than 5 percentage points higher than the random walk model. All but the random walk coefficient are statistically significant. The $R^2$ of our model amounts to 2.44\%, compared to 1.6\% for the analysts, 2.23\% for the regression model and 1.45\% for the random walk. Total assets and the interaction effect of total assets with the surprise variable have both a significant negative effect on the abnormal return for all models (except random walk), which implies that increasing firm size decreases the impact of surprising earnings announcements on the abnormal return. Tobin's q has also a significant negative effect on the abnormal return for all models. However, the interaction effect is not significant. The results for the year dummies are mixed: We get significant positive effects for the years 2015, 2016, 2018, 2019 and 2022 for all models.

\FloatBarrier
\section{Conclusion}
\label{paper2_sec_conclusion}
This study addresses the limitations associated with financial analysts' representation of public knowledge regarding firms' earnings prospects. Previous research has revealed the existence of biases in analysts' estimates due to conflicts of interest, resulting in limited predictive power in explaining stock price reactions to earnings disclosures. Furthermore, analyst coverage remains constrained, both inside and particularly outside the United States. Several subsequent research studies have endeavored to tackle the aforementioned issues, but they themselves exhibit certain weaknesses. These studies employ models that either generate less accurate earnings forecasts compared to financial analysts \citep{hou2012implied, li2014evaluating, hess2019incorporating}, incorporate analyst forecasts in their estimations without resolving the problem of uncovered firms \citep{azevedo2021earnings,ball2018automated}, or rely heavily on data-intensive approaches, thereby introducing selection biases \citep{bradshaw2012re, elend2020earnings}.

To overcome these challenges, this chapter introduces a novel approach by developing a recurrent neural network model for forecasting future firm earnings. Using four decades of quarterly financial information allows the model to learn time dependent patterns in earnings results, which reduces the informational advantage of analysts. Additionally, we avoid the data intensity of previous time-series models where survivorship biases are commonly observed. Our study goes beyond the traditional focus on fiscal-year-end predictions and extends its forecast capabilities to include quarterly earnings, as these are recognized as a crucial source of information that significantly influences stock prices.

The results of this study for annual earnings results indicate that our model surpasses the considered benchmarks, including analyst forecasts, the best cross-sectional regression model of \cite{hess2019incorporating} and the random walk model, in terms of both accuracy and bias. Furthermore, our results demonstrate that the strong predictive power of our model extends to uncovered firms, effectively addressing the issue of low analyst coverage for small or young firms and those operating outside the United States. This outcome signifies the broader applicability and reliability of our model in capturing and forecasting earnings for a diverse range of firms.
In addition, our model exhibits robustness in the face of crises, as indicated by our analysis of the change in predictive power during the Covid-19 crisis. This resilience further underscores the effectiveness and stability of our model in providing accurate earnings predictions even during periods of economic uncertainty.
Notably, when compared to the benchmarks, our model stands out by producing the highest earnings response coefficient. This result signifies that our model's earnings predictions align more closely with market expectations, highlighting its ability to provide precise and reliable insights into firms' future earnings prospects.

Regarding quarterly earnings predictions, our RNN model exhibits lower bias compared to all the benchmarks and more accurate predictions than the regression and the random walk model, although its forecasts are slightly less precise than analyst predictions. Nevertheless, the overall findings remain consistent with the outcomes observed in annual predictions. Given the limited analyst coverage, particularly for quarterly earnings data, the RNN model retains its value as a valuable tool for forecasting uncovered firms. This is especially important considering that the earnings response coefficient of the RNN model is comparable to that of the analysts. Thus, our RNN model serves as a reliable and effective means of generating forecasts for firms that lack analyst coverage, ensuring comprehensive insights for investors and decision-makers.

\newpage
\bibliographystyle{jf}
\bibliography{References.bib}
\newpage

\section*{Appendix}
\addcontentsline{toc}{section}{Appendix}
\subsection*{Performance Per Firm Size On Quarterly Earnings}

\begin{table}
\begin{flushleft}
\caption{Quarterly Performance on Different Firm Sizes (Median Absolute Percentage Difference)}
 \mytablefont{This table presents the median absolute percentage difference (MAPD) between realised and predicted quarterly earnings per share, as defined in section \ref{paper2_sec_mpd_mapd}, partitioned by firm size. We define size deciles using total assets. We compare the RNN model (section \ref{paper2_sec_model}) with analyst forecasts, the cross-sectional regression model of \cite{hess2019incorporating} and the random walk model \citep{bradshaw2012re}, see section \ref{paper2_sec_benchmarks}.}
\label{paper2_tab_firm_size_prediction_accuracyQ}
\mytablefont
\begin{tabularx}{\textwidth}{c *{6}{Y}}
\toprule
\textbf{Size Decile} & \textbf{RNN} & \textbf{Analyst} & \textbf{Regression} & \textbf{Random Walk} & \textbf{Num. Obs.} \\
\midrule
1 & 29.70 \% & 25.16 \% & 52.33 \% & 30.29 \% & 5,033 \\
2 & 30.73 \% & 29.19 \% & 50.18 \% & 33.96 \% & 5,032 \\
3 & 33.68 \% & 32.53 \% & 52.93 \% & 40.23 \% & 5,032 \\
4 & 34.21 \% & 33.87 \% & 53.19 \% & 41.72 \% & 5,033 \\
5 & 33.27 \% & 29.48 \% & 52.17 \% & 40.72 \% & 5,032 \\
6 & 33.13 \% & 28.86 \% & 51.95 \% & 42.66 \% & 5,032 \\
7 & 31.38 \% & 25.70 \% & 49.49 \% & 40.22 \% & 5,033 \\
8 & 29.48 \% & 23.84 \% & 48.45 \% & 38.67 \% & 5,032 \\
9 & 28.42 \% & 22.68 \% & 47.24 \% & 37.68 \% & 5,032 \\
10 & 24.64 \% & 17.91 \% & 46.04 \% & 32.43 \% & 5,033 \\
\bottomrule
\end{tabularx}
\end{flushleft}
\end{table}

The analysis of the model performance with respect to quarterly earnings results reveals similar patterns as the analysis of annual data from section \ref{paper2_sec_firm_size}, as we see in Table \ref{paper2_tab_firm_size_prediction_accuracyQ}. The general pattern of a declining performance in the first three to four deciles and improving performance in the following deciles is still present for all models. For all deciles, analyst forecasts outperform the RNN. However, similar to the annual case, the RNN model performs best compared to the analyst forecasts for the small to medium sized deciles two, three and four, where the gap in performance is around 1 percentage point. For the remaining size deciles, the disparity becomes increasingly pronounced for decile one as well as the higher deciles. Notably, the largest difference is observed in the 10th decile, nearing a margin of approximately 7 percentage points. The RNN outperfroms the other benchmarks for every decile.

\subsection*{Performance Per Industry On Quarterly Earnings}
\begin{table}
\begin{flushleft}
\caption{Quarterly Performance on Different Industries (Median Absolute Percentage Difference)}
 \mytablefont{This table presents the median absolute percentage difference (MAPD) between realised and predicted quarterly earnings per share, as defined in section \ref{paper2_sec_mpd_mapd}, partitioned by industry. We use the economic sector class of The Refinitiv Business Classifications (TRBC) for industry classification. We compare the RNN model (section \ref{paper2_sec_model}) with analyst forecasts, the cross-sectional regression model of \cite{hess2019incorporating} and the random walk model \citep{bradshaw2012re}, see section \ref{paper2_sec_benchmarks}.}
\label{paper2_tab_industry_prediction_accuracyQ}
\mytablefont
\begin{tabularx}{\textwidth}{c *{6}{Y}}
\toprule
\textbf{Industry} & \textbf{RNN} & \textbf{Analyst} & \textbf{Regression} & \textbf{Random Walk} & \textbf{Num. Obs.} \\
\midrule
Academic \& Educational Services & 27.43 \% & 17.03 \% & 48.08 \% & 45.33 \% & 222 \\
Basic Materials & 39.93 \% & 25.72 \% & 54.78 \% & 46.87 \% & 2,966 \\
Consumer Cyclicals & 33.66 \% & 24.49 \% & 50.98 \% & 47.96 \% & 9,310 \\
Consumer Non-Cyclicals & 25.41 \% & 20.04 \% & 43.50 \% & 34.45 \% & 2,903 \\
Energy & 70.72 \% & 54.72 \% & 81.76 \% & 69.19 \% & 2,462 \\
Financials & 26.27 \% & 18.86 \% & 49.64 \% & 34.03 \% & 753 \\
Healthcare & 25.26 \% & 22.16 \% & 47.06 \% & 24.99 \% & 10,360 \\
Industrials & 28.29 \% & 21.26 \% & 47.22 \% & 34.50 \% & 8,873 \\
Other & 65.26 \% & 24.28 \% & 40.08 \% & 58.63 \% & 40 \\
Real Estate & 35.46 \% & 25.20 \% & 52.57 \% & 45.39 \% & 308 \\
Technology & 32.74 \% & 46.70 \% & 53.71 \% & 36.21 \% & 10,237 \\
Utilities & 23.33 \% & 14.26 \% & 49.08 \% & 53.81 \% & 1,890 \\
\bottomrule
\end{tabularx}
\end{flushleft}
\end{table}

Table \ref{paper2_tab_industry_prediction_accuracyQ} presents the performance evaluation for quarterly earnings predictions of the RNN model and all benchmarks segmented by industry. We see that our model outperforms the random walk and the regression model for almost every industry. However, comparing to quarterly analyst forecasts, the results indicate that the analyst forecasts outperform the RNN model for all industries, with differences varying between 16 percentage points for the energy sector and 3.1 percentage points in the healthcare sector. The only exception is the technology sector, where our model performs more than 14 percentage points better than the analyst. This sector is with 10,237 observations the second largest industry in the sample. 

\subsection*{Sign Prediction On Quarterly Earnings}
\begin{table}
\begin{flushleft}
\caption{Sign Prediction (Quarterly)}
 \mytablefont{This table presents the results for the sign prediction analysis for quarterly data, as defined in section \ref{paper2_sec_sign_pred_def}, for the RNN model (section \ref{paper2_sec_model}) and all benchmarks. This analysis examines the models ability to predict the direction of an earnings change from the previous to the actual period. We define three classes: positive (percentage change > 5\%), negative (percentage change < -5\%) and neutral (|percentage change| < 5\%). The benchmarks are analysts forecasts, the cross-sectional regression model of \cite{hess2019incorporating} and the random walk model \citep{bradshaw2012re}, see section \ref{paper2_sec_benchmarks}. We calculate macro accuracy, precision, recall and F1 for all models, as defined in \cite{sokolova2009systematic}.}
\label{paper2_tab_sign_predictionQ}
\mytablefont
\begin{tabularx}{\textwidth}{c *{5}{Y}}
\toprule
 & \textbf{RNN} & \textbf{Analyst} & \textbf{Random Walk} & \textbf{Regression} \\
\midrule
Accuracy & 76.71 \% & 78.99 \% & 77.00 \% & 68.56 \% \\
Precision & 54.04 \% & 58.07 \% & 54.55 \% & 43.06 \% \\
Recall & 54.51 \% & 59.23 \% & 53.84 \% & 40.72 \% \\
F1 & 53.83 \% & 58.53 \% & 53.91 \% & 38.48 \% \\
Num. Obs. & \multicolumn{4}{c}{42,368} \\
\bottomrule
\end{tabularx}
\end{flushleft}
\end{table}

Table \ref{paper2_tab_sign_predictionQ} shows the results for the sign prediction for quarterly earning results of the RNN model and all benchmarks. We see that the RNN model is outperformed by the analyst forecasts with respect to accuracy, precision recall and F1. The RNN model performs similar to the random walk model. Only the regression model performs worse than the other models. 

\subsection*{Performance Covered vs. Uncovered Firms}
\begin{table}
\begin{flushleft}
\caption{Performance Analyst Coverage vs. No Analyst Coverage}
 \mytablefont{This table compares the median absolute percentage difference (MAPD) and the median percentage difference (MPD), as defined in section \ref{paper2_sec_mpd_mapd}, for analyst-covered and not analyst-covered firms. We compute these measures for the RNN model (section \ref{paper2_sec_model}), analyst forecasts, the cross-sectional regression model of \cite{hess2019incorporating} and the random walk model \citep{bradshaw2012re}, see section \ref{paper2_sec_benchmarks}. We report the results for annual (Panel A) and quarterly (Panel B) predictions. For all models, we test whether their median value for covered firms is significantly different to their median value for uncovered firms using the Mann-Whitney U test \citep{mann1947test}.}
\label{paper2_tab_covered_uncovered}
\mytablefont
\begin{tabularx}{\textwidth}{c *{7}{Y}}
\toprule
 & \multicolumn{2}{c}{\textbf{Covered}} & \multicolumn{2}{c}{\textbf{Uncovered}} & \multicolumn{2}{c}{\textbf{P Value}} \\
 & MAPD & MPD & MAPD & MPD & MAPD & MPD \\
\midrule
   \multicolumn{7}{c}{\textbf{Panel A: Annual}} \\
       \midrule
RNN & 19.56 \% & 1.01 \% & 28.38 \% & 0.49 \% & 0.00 \% & 1.19 \% \\
Regression & 28.17 \% & 5.86 \% & 37.14 \% & 4.97 \% & 0.00 \% & 3.57 \% \\
Random Walk & 34.52 \% & 5.94 \% & 46.05 \% & 0.29 \% & 0.00 \% & 0.00 \% \\
Num. Obs. & \multicolumn{2}{c}{52,445} & \multicolumn{2}{c}{8,823} & &  \\
\midrule
  \multicolumn{7}{c}{\textbf{Panel B: Quarterly}} \\
       \midrule
RNN & 31.22 \% & 1.08 \% & 43.38 \% & -0.27 \% & 0.00 \% & 1.18 \% \\
Regression & 50.33 \% & 11.95 \% & 58.09 \% & 7.08 \% & 0.00 \% & 1.45 \% \\
Random Walk & 37.79 \% & 2.03 \% & 51.72 \% & -0.00 \% & 0.00 \% & 0.04 \% \\
Num. Obs. & \multicolumn{2}{c}{52,199} & \multicolumn{2}{c}{8,747} & &  \\
\bottomrule
\end{tabularx}
\end{flushleft}
\end{table}

Table \ref{paper2_tab_covered_uncovered} investigates the difference in performance of our model and the remaining benchmarks in both cases, analyst coverage and no analyst coverage. For annual predictions, we see that the MAPD of our model increases about 9 percentage points if we consider data that is not covered by analysts. A similar decrease in performance can be observed for the benchmarks, too. Regarding the MPD, we observe the opposite pattern. For all models, the predictions get less biased if the data is not covered by analysts. All of these effects are statistically significant on the 5\% level.

Looking at the quarterly predictions, we see the same pattern. The MAPD of all models increases between 8 and 14 percentage points if the data is not covered. For all models, the bias decreases for uncovered data. All these effects are statistically significant at the 1\% level.

\begin{table}
\begin{flushleft}
\caption{Performance Analyst Coverage vs. Similar No Analyst Coverage (Quarterly)}
 \mytablefont{This table presents the median absolute percentage difference (MAPD), as defined in section \ref{paper2_sec_mpd_mapd}, of the quarterly RNN model EPS predictions (see section \ref{paper2_sec_model}) and realised EPS values for not analyst-covered firms and similar analyst-covered firms. Two firms are similar if they are in the same industry and if the eucledian distance between their arrays consisting of standarized values of total assets and Tobin's q is smaller than 0.01. The sample consists of all non-covered firm observations for which there is at least one similar covered firm observation. If there is more than one similar firm observation, we take the most similar one. We test whether the median value for non-covered firms is significantly different to the median value for similar covered firms using the Wilcoxon rank test \citep{wilcoxon1992individual}.}
\label{paper2_tab_covered_similar_uncoveredQ}
\mytablefont
\begin{tabularx}{\textwidth}{m{2cm} *{4}{Y}}
\toprule
 & Count & \shortstack{MAPD \\ Covered} & \shortstack{MAPD \\ Uncovered} & P Value \\
\midrule
Academic \& Educational Services & 70 & 29.0 \% & 57.1 \% & 0.7 \% \\
Basic Materials & 821 & 46.1 \% & 44.7 \% & 74.4 \% \\
Consumer Cyclicals & 4,207 & 41.7 \% & 55.1 \% & 0.0 \% \\
Consumer Non-Cyclicals & 517 & 49.5 \% & 30.6 \% & 0.0 \% \\
Energy & 860 & 72.8 \% & 81.2 \% & 23.7 \% \\
Financials & 135 & 51.0 \% & 47.6 \% & 93.7 \% \\
Healthcare & 6,804 & 24.1 \% & 26.0 \% & 0.0 \% \\
Industrials & 4,115 & 35.9 \% & 29.2 \% & 0.0 \% \\
Other & 5 & 108.7 \% & 81.3 \% & 81.2 \% \\
Real Estate & 40 & 26.5 \% & 62.9 \% & 0.3 \% \\
Technology & 4,801 & 40.6 \% & 44.8 \% & 0.0 \% \\
Utilities & 157 & 19.5 \% & 28.2 \% & 0.9 \% \\
 \midrule
Total & 22,532 & 34.7 \% & 37.6 \% & 0.0 \% \\
\bottomrule
\end{tabularx}
\end{flushleft}
\end{table}

Table \ref{paper2_tab_covered_similar_uncoveredQ} presents the results of the comparison between the RNN's predictive performance for covered and similar uncovered firms, based on quarterly earnings results. The exact procedure is described in section \ref{paper2_sec_covered_uncovered}. We see that the RNN's MAPD of 37.6\% is comparable to the 34.7\% for covered firms. Especially if we compare this to the naive comparison of covered and uncovered firms, without considering firm similarities. In this case, the difference in the models predictive performance on covered and uncovered firms is more than 12 percentage points. Regarding the results on industry level, we see that the model achieves compatible results for similar uncovered firms from most of the considered industries. 

\subsection*{Performance Pre vs. During Covid On Quarterly Earnings}
\begin{table}
\begin{flushleft}
\caption{Performance Pre vs. During COVID 19 (Quarterly)}
 \mytablefont{This table compares the median absolute percentage difference (MAPD) and the median percentage difference (MPD), as defined in section \ref{paper2_sec_mpd_mapd}, for quarterly earnings reports before and during the Covid pandemic. We compute these measures for the RNN model (section \ref{paper2_sec_model}), analyst forecasts, the cross-sectional regression model of \cite{hess2019incorporating} and the random walk model \citep{bradshaw2012re}, see section \ref{paper2_sec_benchmarks}. We set February 18, 2020 as the starting point for the pandemic. For all models, we test whether their median value for observations prior to the pandemic is significantly different to their median value for observations during the pandemic using the Mann-Whitney U test \citep{mann1947test}.}
\label{paper2_tab_covid_predictionQ}
\mytablefont
\begin{tabularx}{\textwidth}{c *{7}{Y}}
\toprule
 & \multicolumn{2}{c}{\textbf{Pre-Covid}} & \multicolumn{2}{c}{\textbf{During-Covid}} & \multicolumn{2}{c}{\textbf{P Value}} \\
 & MAPD & MPD & MAPD & MPD & MAPD & MPD \\
\midrule
RNN & 28.66 \% & 1.05 \% & 36.50 \% & 2.58 \% & 0.00 \% & 0.00 \% \\
Analyst & 25.11 \% & -7.55 \% & 30.81 \% & -5.83 \% & 0.00 \% & 0.00 \% \\
Regression & 49.55 \% & 13.93 \% & 52.30 \% & 4.90 \% & 0.00 \% & 0.00 \% \\
Random Walk & 37.26 \% & 2.99 \% & 38.25 \% & 0.87 \% & 26.91 \% & 88.74 \% \\
Num. Obs. & \multicolumn{2}{c}{36,714} & \multicolumn{2}{c}{13,610} & &  \\
\bottomrule
\end{tabularx}
\end{flushleft}
\end{table}

Table \ref{paper2_tab_covid_predictionQ} shows the results for the comparison of the predictive performance of RNN and benchmarks before and during the Covid crisis for quarterly earnings results. In contrast to results for annual earnings in section \ref{paper2_sec_covid}, the RNN model reacts more sensitive to the Covid crisis, as the MAPD increases from 28.66\% before Covid to 36.5\% during Covid. The MAPD of the analyst forecasts only increases about 5 percentage points. The other benchmarks are even more robust to the Covid crisis. The same holds for the bias of the RNN model. The MPD is increasing during Covid for the RNN model, while it is decreasing for all benchmarks. These differences are all statistically significant at the 1\% level, except for the random walk model.

\subsection*{Earnings Response Coefficient On Quarterly Earnings}
\begin{table}
\begin{flushleft}
\caption{Earnings Response Coefficient (Quarterly)}
 \mytablefont{This table presents the quarterly results for the earnings response regression of section \ref{paper2_sec_ERC}. We fit a year- and firm-fixed effects model with clustered standard errors. Earnings surprise is defined as $\frac{EPS-\widehat{EPS}}{\widehat{EPS}}$ with $EPS$ being the realized earnings per share and $\widehat{EPS}$ is its prediction.}
\label{paper2_tab_ercQ}
\mytablefont
\begin{tabularx}{\textwidth}{c *{5}{Y}}
\toprule
 & \textbf{RNN} & \textbf{Analyst} & \textbf{Regression} & \textbf{Random Walk} \\
\midrule
surprise (\%) & 5.17*** & 5.26*** & 2.70*** & 2.34*** \\
lnTA (\%) & -2.41*** & -2.31*** & -2.56*** & -2.43*** \\
lnAssets x surprise (\%) & -0.41*** & -0.46*** & -0.19*** & -0.19*** \\
Tobin's q (\%) & -0.35*** & -0.34*** & -0.36*** & -0.34*** \\
Tobin's q x surprise (\%) & -0.06*** & -0.03 & -0.04*** & -0.01 \\
Q1 (\%) & 0.00 & -0.30 & 0.15 & -0.10 \\
Q2 (\%) & 0.08 & -0.13 & 0.20 & -0.22 \\
Q3 (\%) & 0.09 & -0.05 & 0.18 & -0.04 \\
2015 (\%) & -0.10 & -0.16 & -0.13 & -0.17 \\
2016 (\%) & 0.40*** & 0.37*** & 0.34*** & 0.29** \\
2017 (\%) & -0.21 & -0.28* & -0.24 & -0.28* \\
2018 (\%) & 0.76*** & 0.73*** & 0.82*** & 0.75*** \\
2019 (\%) & 0.81*** & 0.77*** & 0.89*** & 0.71*** \\
2020 (\%) & 1.16*** & 0.91*** & 1.20*** & 0.91*** \\
2021 (\%) & 0.97*** & 0.85*** & 1.04*** & 1.02*** \\
2022 (\%) & 2.14*** & 1.98*** & 2.46*** & 2.18*** \\
$R^2$ (\%) & 3.50 & 3.11 & 2.44 & 2.22 \\
Num. Obs. & 50,226 & 50,226 & 50,226 & 50,226 \\
\bottomrule
\end{tabularx}
\end{flushleft}
\end{table}

Table \ref{paper2_tab_ercQ} contains the results of the earnings response analysis, as described in section \ref{paper2_sec_ERC}, for quarterly earnings. We see that the results of the RNN model and the analyst forecasts are quite similar. The analysts ERC is slightly higher compared to our model (5.26\% and 5.17\%, respectively). However, the $R^2$ of the RNN is with 3.5\% higher than the analysts' $R^2$ of 3.11\%. The other benchmarks perform worse, around 3 percentage points worse with respect to ERC and more than 1 percentage point worse in terms of $R^2$. The effects for total assets, Tobin's q and the respective interaction terms are similar to the annual case, with the only exception that the interaction of Tobin's q is statistically significant for our model and the regression model. The quarter dummies are insignificant for all models. The year dummies of 2016 and 2018-2022 are all significant and positive for all models. 2015 is insignificant for every model. 2017 is only significant for the analyst forecasts and the random walk. 

\subsection*{Performance Per Year}
\begin{table}
\begin{flushleft}
\caption{Performance Per Year}
 \mytablefont{This table presents the median absolute percentage difference (MAPD) and the median percentage difference (MPD) between realised and predicted earnings per share, as defined in section \ref{paper2_sec_mpd_mapd}, partitioned by fiscal year. We compare the RNN model (section \ref{paper2_sec_model}) with analysts forecasts, the cross-sectional regression model of \cite{hess2019incorporating} and the random walk model \citep{bradshaw2012re}, see section \ref{paper2_sec_benchmarks}. We report the results for annual (Panel A) and quarterly (Panel B) predictions.}
\label{paper2_tab_yearly_performance}
\mytablefont
\begin{tabularx}{\textwidth}{ *{10}{Y}}
\toprule
 & \multicolumn{2}{c}{\textbf{RNN}} & \multicolumn{2}{c}{\textbf{Analyst}} & \multicolumn{2}{c}{\textbf{Regression}} & \multicolumn{2}{c}{\textbf{Random Walk}} & \textbf{Num.} \\
 \textbf{Year} & MAPD & MPD & MAPD & MPD & MAPD & MPD & MAPD & MPD &\textbf{Obs.} \\
\midrule
   \multicolumn{10}{c}{\textbf{Panel A: Annual}} \\
       \midrule
2014 & 15.2 \% & 1.5 \% & 22.1 \% & -12.4 \% & 23.9 \% & 10.9 \% & 28.3 \% & 6.5 \% & 4,923 \\
2015 & 17.1 \% & -0.7 \% & 24.0 \% & -16.3 \% & 25.7 \% & 6.2 \% & 29.2 \% & 3.1 \% & 6,405 \\
2016 & 16.2 \% & 0.3 \% & 22.3 \% & -15.5 \% & 26.1 \% & 9.5 \% & 28.8 \% & 6.4 \% & 6,298 \\
2017 & 26.4 \% & 3.3 \% & 31.1 \% & -9.7 \% & 33.6 \% & 10.9 \% & 37.0 \% & 11.4 \% & 6,218 \\
2018 & 20.6 \% & 2.0 \% & 24.0 \% & -11.3 \% & 31.0 \% & 6.4 \% & 37.5 \% & 10.8 \% & 6,465 \\
2019 & 17.4 \% & -0.6 \% & 25.0 \% & -18.3 \% & 25.8 \% & -1.7 \% & 29.8 \% & -0.0 \% & 6,494 \\
2020 & 23.0 \% & 0.4 \% & 32.6 \% & -17.2 \% & 29.0 \% & 0.6 \% & 39.1 \% & -1.1 \% & 6,464 \\
2021 & 20.1 \% & 3.1 \% & 24.4 \% & -7.8 \% & 30.9 \% & 2.7 \% & 43.7 \% & 19.8 \% & 7,224 \\
\midrule
   \multicolumn{10}{c}{\textbf{Panel B: Quarterly}} \\
       \midrule
2014 & 26.3 \% & 2.1 \% & 24.0 \% & -6.9 \% & 50.2 \% & 23.9 \% & 33.9 \% & 6.1 \% & 4,869 \\
2015 & 26.8 \% & -1.9 \% & 23.8 \% & -9.0 \% & 48.4 \% & 11.7 \% & 35.6 \% & 0.2 \% & 6,318 \\
2016 & 26.6 \% & 0.1 \% & 23.6 \% & -8.3 \% & 47.6 \% & 15.5 \% & 34.6 \% & 3.1 \% & 6,235 \\
2017 & 32.5 \% & 1.9 \% & 28.6 \% & -5.7 \% & 52.0 \% & 16.3 \% & 41.7 \% & 5.4 \% & 6,298 \\
2018 & 32.0 \% & 5.9 \% & 25.8 \% & -5.7 \% & 53.9 \% & 15.9 \% & 42.5 \% & 2.8 \% & 6,303 \\
2019 & 28.8 \% & -1.1 \% & 25.9 \% & -10.7 \% & 48.0 \% & 1.0 \% & 36.4 \% & -0.0 \% & 6,419 \\
2020 & 39.7 \% & 0.2 \% & 33.7 \% & -7.3 \% & 54.4 \% & -3.6 \% & 44.2 \% & 1.6 \% & 6,420 \\
2021 & 33.3 \% & 5.7 \% & 27.8 \% & -3.6 \% & 49.1 \% & 13.3 \% & 33.4 \% & 1.5 \% & 7,379 \\
\bottomrule
\end{tabularx}
\end{flushleft}
\end{table}

In Table \ref{paper2_tab_yearly_performance} we investigate the predictive performance of the RNN model and the benchmarks for each year in the test set separately. For annual predictions, we see that our model outperforms the analyst forecasts in every year with margins between 4 and 7 percentage points. The same holds, with even higher margins, for the other benchmarks. There is a general trend of a slightly decreasing performance with increasing years for the estimated models, i.e.~the RNN and the regression model. This observation signifies that the model acquires temporal patterns during the training phase, which progressively diminish or undergo alterations in the test set as the temporal distance between the observation and the training dataset increases. Additionally, we see that for all years, the RNN model is less biased than all benchmarks.

Repeating the same analysis for quarterly earnings results, we deduce that the RNN performs worse than the analyst forecasts in all years with respect to MAPD. However, it outperforms all other benchmarks. The trend of a slightly decreasing performance with respect to MAPD exists for quarterly earnings predictions too.

\end{document}